\documentclass[%
reprint,
superscriptaddress,
 amsmath,amssymb,
 aps,
prb,
nofootinbib
]{revtex4-2}

\usepackage{physics}
\usepackage[dvipsnames]{xcolor} 
\usepackage{graphicx}
\usepackage{dcolumn}
\usepackage{bm}
\usepackage{xcolor, soul}
\sethlcolor{yellow}
\usepackage[makeroom]{cancel}

\makeatletter
\def\@fnsymbol#1{%
   \ifcase#1\or
   \TextOrMath \textdagger \dagger\or
   \TextOrMath \textasteriskcentered *\or
   \TextOrMath \textdaggerdbl \ddagger \or
   \TextOrMath \textsection  \mathsection\or
   \TextOrMath \textparagraph \mathparagraph\or
   \TextOrMath \textbardbl \|\or
   \TextOrMath {\textdagger\textdagger}{\dagger\dagger}\or
   \TextOrMath {\textdaggerdbl\textdaggerdbl}{\ddagger\ddagger}\else
   \@ctrerr \fi
}
\makeatother

\usepackage[T1]{fontenc}

\begin{document}

\title{Application of the exact-factorization density-functional perturbation approach\\ to pentacene crystal and monolayer MoS$_2$}

\author{Rachel Steinitz-Eliyahu}
\thanks{These authors contributed equally to this work.}
\affiliation{Department of Molecular Chemistry and Materials Science, Weizmann Institute of Science, Rehovot 7610001, Israel}

\author{Galit Cohen}
\thanks{These authors contributed equally to this work.}
\affiliation{Department of Molecular Chemistry and Materials Science, Weizmann Institute of Science, Rehovot 7610001, Israel}
\affiliation{Oden Institute for Computational Engineering and Sciences, The University of Texas at Austin, Austin, TX 78712}

\author{Guy Vosco}
\affiliation{Department of Molecular Chemistry and Materials Science, Weizmann Institute of Science, Rehovot 7610001, Israel}

\author{E.K.U. Gross}
\affiliation{Quantum Dynamics Laboratory, Tsientang Institute of Advanced Study, Hangzhou, Zhejiang, 310024, P.R. China}
\affiliation{Fritz Haber Center for Molecular Dynamics, Institute of Chemistry, The Hebrew University of Jerusalem, Jerusalem 91904, Israel}

\author{Ryan Requist}
\affiliation{Division of Theoretical Physics, Ru\dj er Bo\v{s}kovi\'{c} Institute, Zagreb 10000, Croatia}

\author{Sivan Refaely-Abramson}
\affiliation{Department of Molecular Chemistry and Materials Science, Weizmann Institute of Science, Rehovot 7610001, Israel}

\begin{abstract}
Non-adiabatic effects arising from electron–phonon interactions are often neglected within the Born–Oppenheimer (BO) approximation, which assumes that electronic states adjust instantaneously to nuclear motion. The exact factorization (EF) formalism provides a rigorous framework for treating such effects beyond the adiabatic regime and has recently been adapted to density functional theory (DFT) in the harmonic limit. Building on these foundations, we previously introduced an EF-based perturbative scheme, the EF density-functional perturbation theory (EF-DFPT), that enables the computation of phonon-driven non-adiabatic (NA) corrections to Kohn–Sham (KS) electronic states, up to second order in nuclear displacements.
Here, we present the first implementation and application of EF-DFPT to extended periodic materials, focusing on its impact on experimentally relevant observables. Using the pentacene molecular crystal and monolayer MoS$_2$ as representative soft- and stiff-mode systems, respectively, we demonstrate how NA electron–phonon interactions modify the static dielectric response. We show that these modifications originate from the combined effect of NA phonon-dressed electronic wavefunctions and second-order NA energy renormalizations. The resulting behavior is strongly material dependent: NA effects are negligible in monolayer MoS$_2$, whereas in pentacene they lead to pronounced long-range screening effects associated with soft vibrational modes and enhanced electron–phonon coupling.
Our results establish EF-DFPT as a practical \textit{ab initio} framework for describing non-adiabatic electron–phonon interactions in solids and lay the groundwork for future EF-based approaches capable of capturing fully non-adiabatic phenomena in materials.
\end{abstract}

\maketitle
The coupling between electrons and phonons plays a central role in determining the structural, electronic, and dynamical properties of materials. Electron–phonon (el–ph) interactions shape ground-state and excited-state properties associated with charge transport, energy conversion and spectroscopy~\cite{giustino2010electron,giustino2008small,eiguren2003electron,valla1999many,lafuente2022ab,lafuente2022unified,sio2023polarons,perfetto2023real,stefanucci2023and}.  Substantial advances in both theory development and computational capabilities allow the calculation of these interactions using predictive \textit{ab initio} approaches based on density functional perturbation theory (DFPT)~\cite{gonze1995adiabatic,baroni2001phonons,giustino2017electron}. These typically rely on calculations of the adiabatic response of the Kohn-Sham (KS) potential within the Born-Oppenheimer (BO) approximation~\cite{BO,ziman1960electrons}, which assumes that electrons respond instantaneously to nuclear displacements and thus captures only adiabatic effects. This approximation proves to be effective for a vast variety of physical phenomena; however, it fails by construction in capturing non-adiabatic (NA) effects~\cite{girotto2023ultrafast,malhado2014non,yarkony2012nonadiabatic,yonehara2012fundamental,miglio2020predominance} originating from changes in the electron distribution due to nuclear movement. Despite a number of recent efforts to capture important pieces of non-adiabatic effects from \textit{ab initio}~\cite{BergesPRX2023,calandra2010adiabatic,LazzeriPRL2006,ponce2015temperature,CarusoPRL2017,verdi2017origin,girotto2023ultrafast}, a fully general, predictive and computationally feasible description of el-ph interactions beyond the BO approximation is still a challenge.
Non-adiabatic effects are particularly important in systems where electron–phonon coupling is strong, such as materials hosting polarons- quasiparticles formed through the mutual interaction of electrons and lattice degrees of freedom~\cite{frohlich1954electrons,lafuente2022ab,lafuente2022unified,sio2023polarons,dai2025polarons}. Since the nature of the dielectric environment depends on the interactions between electrons and nuclei, polaronic effects can be observed through the dielectric response in materials~\cite{zhang2022ab, Zacharias_temp}, with direct impact on the optoelectronic properties and materials functionality~\cite{franchini2021polarons}. Materials with different atomic structure and dimensionality respond differently to an external electric field, as determined by their polarizability. Organic molecular solids, such as the pentacene crystal which is broadly used in organic electronics and energy capture applications, have soft phonon modes that promote strong electron–phonon coupling and polaronic effects \cite{kronik2016excited,cohen2024phonon,refaely2013gap}. On the other limit, monolayer MoS$_2$, a transition metal dichalcogenide semiconductor widely studied in optoelectronic science and spectroscopy, is a hard-mode system with weak electron-phonon coupling~\cite{li2021exciton,jung2024holstein}. Predictive, \textit{ab initio} computations of NA effects and the associated changes in the electronic polarizability can shed light on the relation between atomistic structure and dynamics to dielectric and electronic phenomena, in these materials and in general. 

The exact factorization (EF)~\cite{abedi2010exact,gidopoulos2014electronic} method employs the factorization of the many-electron, many-nucleus wavefunction. This formalism is formally exact, i.e.~retains all non-adiabatic effects, and was employed to deduce a density functional framework for the complete system of electrons and nuclei~\cite{requist2016exact, li2018density}. For the case of small-amplitude nuclear motion, this can be cast into a density functional theory (DFT) for electrons and phonons, as recently derived by Requist {\it et al.}~\cite{RPG2019}. In this approach the Kohn-Sham electrons are ''dressed'' by
non-adiabatic phonon-induced interactions that appear through a non-adiabatic correction to the Kohn-Sham potential. The self-consistent density obtained is the conditional probability of finding an electron at a position $\mathbf{r}$, given that the collection of nuclei resides at a certain point in nuclear configuration space.
So far, this formalism has not been implemented to treat realistic materials.  
In our recent work~\cite{cohen2025nonadiabaticity}, we derived a DFPT-based methodology within the EF framework, presenting non-adiabatic effects in the electronic equation that are introduced through an additive non-local DFT potential, which is added on top of the adiabatic KS potential. 
The nuclear displacements are then treated perturbatively, introducing first and second order adiabatic and non-adiabatic perturbative potentials. This leads to both adiabatic and non-adiabatic corrections in the electronic states due to phonon dressing. 

In this work,  we present the first \textit{ab initio} calculation of non-adiabatic corrections within the EF-DFPT framework for extended semiconducting materials, explicitly accounting for electron–lattice coupling beyond the adiabatic approximation in a Kohn–Sham-like formulation. We demonstrate our approach for two limiting cases of reduced-dimensional semiconductors, the stiff-mode monolayer transition metal dichalcogenide MoS$_2$, and the soft-mode pentacene molecular crystal. Our central objective is to quantify how NA electron–phonon interactions manifest in experimentally relevant observables. To this end, we compute first- and second-order NA corrections to electronic wavefunctions and energies, explicitly including multi-band transitions, and analyze how these corrections propagate into the dielectric response. We show that NA effects are negligible in monolayer MoS$_2$, reaffirming the validity of the Born–Oppenheimer approximation, while they are substantial in pentacene, where soft phonon modes and reduced screening enhance long-range electron–phonon interactions and polaronic behavior. Our study thus constitutes a key first step toward a fully self-consistent, DFT-based EF treatment of non-adiabatic effects in materials.

The EF-DFPT formalism starts with a KS-like electronic Hamiltonian~\cite{RPG2019,cohen2025nonadiabaticity},
\begin{equation}\label{eq:EF_elec_eq}
    \begin{split}
        \Big[\frac{\hat{p}^2}{2m}+v_{en}(\mathbf{r},U)+v_{hxc}(\mathbf{r},U)+v_{geo}&(\mathbf{r},U)\Big] \psi_{n\mathbf{k}U}(\mathbf{r})\\
        &=\varepsilon_{n\mathbf{k}U}\psi_{n\mathbf{k}U}(\mathbf{r})\;.
    \end{split}
\end{equation}
Here $\textbf{r}$ denotes the electronic coordinates and $U = \{U_{\textbf{q}\lambda}\}$ the set of harmonic phonon coordinates, where $\mathbf{q}$ and $\lambda$ label the phonon momentum and mode, respectively.  $\psi_{n\mathbf{k}U}(\mathbf{r})$ stands for conditional single particle orbitals in the generalized Kohn-Sham (GKS) \cite{kummel2008orbital} approach with
$n$ and $\mathbf{k}$ denoting the electronic band index and crystal momentum.
The first three terms in Eq.~\eqref{eq:EF_elec_eq} correspond to the electronic kinetic energy, the electron–lattice interaction potential $\hat{v}_{en}$, and the Hartree-exchange-correlation (HXC) potential $\hat{v}_{hxc}$. Non-adiabatic effects enter through the additional potential, $\hat{v}_{geo}$, derived within the exact factorization formalism, and explicitly defined in ref ~\cite{cohen2025nonadiabaticity,RPG2019}. To solve the KS-like equation, we apply perturbation theory and expand the Hamiltonian up to second order in the nuclear displacements $U_{\textbf{q}\lambda}$. The adiabatic perturbation follows standard DFPT \cite{gonze1995adiabatic,baroni2001phonons, giustino2017electron} and is given by
\begin{equation}\label{eq:v_KS1}
    \mel{\psi^{(0)}_{2}}{\hat{v}_{s}^{(1)}}{\psi^{(0)}_{1}}\equiv \sum_{\lambda} g_{n_2,n_1,\lambda}(\mathbf{k_1},\mathbf{q})\frac{U_{\mathbf{q}\lambda}}{L_{\mathbf{q}\lambda}} \;,
\end{equation}
where the superscript indicates the order of the perturbation, and the subscript, $1\equiv n_1,\mathbf{k_1}$ and $2\equiv n_2,\mathbf{k_2}$. Momentum conservation enforces $\mathbf{k_2} = \mathbf{k_1}+\mathbf{q}$. The perturbation $\hat{v}_{s}^{(1)}$ is comprised of $\hat{v}_{en}^{(1)}+ \hat{v}_{hxc}^{(1)}$. The quantity $L_{\mathbf{q}\lambda} = \sqrt{\frac{\hbar}{2M\omega_{\mathbf{q}\lambda}}}$ is the zero-point motion amplitude, and $g_{n_2,n_1,\lambda}(\mathbf{k_1},\mathbf{q})$ denotes the first-order electron–phonon coupling matrix element.
Within the EF-DFPT framework, the first-order non-adiabatic correction appears as an additive potential to the adiabatic DFPT term,
\begin{equation} \label{eq:v_geo1}
\begin{split}
    &\mel{\psi_{2}^{(0)}}{\hat{v}_{geo}^{(1)}}{\psi_{1}^{(0)}} =\sum_\lambda \hbar\omega_{\mathbf{q}\lambda} U_{\mathbf{q}\lambda} \braket{\psi_{2}}{\pdv{\psi^{(1)}_{1}}{U_{\mathbf{q}\lambda}}}\\
    &= \sum_\lambda g_{n_2,n_1,\lambda}(\mathbf{k_1},\mathbf{q})\frac{U_{\mathbf{q}\lambda}}{L_{\mathbf{q}\lambda}}\frac{\hbar\omega_{\mathbf{q}\lambda}}{\epsilon^{(0)}_{1}-\epsilon^{(0)}_{2}\pm\hbar\omega_{\mathbf{q}\lambda}+i\eta}\;,
\end{split}
\end{equation}
This term mixes electronic states through phonon emission and absorption processes and is therefore off-diagonal in the KS basis. Here after, we refer to non-adiabatic as the corrections that come from using the $\hat{v}_{geo}$ additive term, i.e using $\hat{v}_{s}+\hat{v}_{geo}$ In the present work we consider zero-point motion, $ \langle \hat{U}_{\mathbf{q}\lambda} \rangle = L_{\mathbf{q}\lambda}$; at finite temperature the ratio $\langle \hat{U}_{\mathbf{q}\lambda} \rangle/L_{\mathbf{q}\lambda}>1$ ~\cite{Zacharias_temp} enhances non-adiabatic effects (see SI). Since we consider systems at zero temperature, only the phonon emission case appears hereafter.

The polarizability of a solid is not a purely electronic quantity but it is renormalized by interactions between the electrons and their nuclear environment. When the latter consists of harmonic lattice vibrations, the electron–phonon coupling modifies both the electronic wavefunctions and energies, and therefore the dielectric response.
The dielectric response depends explicitly on two EF-DFPT ingredients incorporated within the polarizability: (i) NA phonon-dressed electronic wavefunctions, and (ii) NA corrections to the electronic energies entering the polarizability. Within the EF-DFPT framework, these effects are incorporated through perturbative corrections to the wavefunctions and energies, which we explicitly calculate here up to first order in the wavefunctions and second-order in the electronic energies. In the following we analyze these two ingredients in detail, in order to better understand the origin of the change in the dielectric response upon the consideration of NA electron-phonon interactions.

\begin{figure}[htbp]
    \centering
    \includegraphics[width=\linewidth]{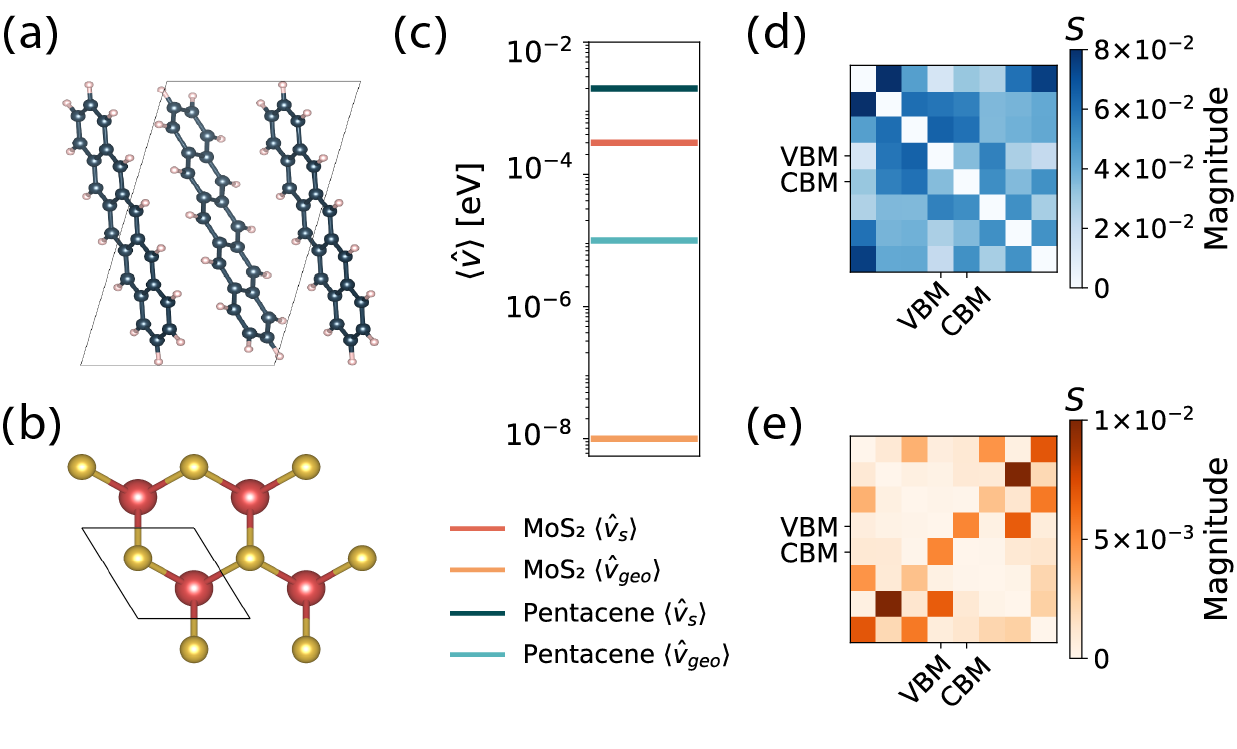}
    \caption{(a),(b) Crystal structures of the computed pentacene and MoS$_2$ systems, respectively. (c) Averaged values of the adiabatic electron–phonon potential and the EF non-adiabatic electron–phonon potential. (d,e)  Overlap matrix elements of the KS orbitals of the $C$ point of pentacene and the $K$ point of MoS$_2$. }
    \label{fig:1}
\end{figure}

Fig.~\ref{fig:1}(a,b) presents the structures of pentacene and MoS$_2$, the characteristic systems examined in this work (for full computational details, see SI).
Fig.~\ref{fig:1}(c) shows the computed average values of the adiabatic and NA perturbations, as given in equations \eqref{eq:v_KS1},\eqref{eq:v_geo1}, respectively, for the pentacene crystal and MoS$_2$ (further details are given in Fig.~S2 in the SI). While in both materials the NA corrections are small, as expected, in pentacene they are about two orders of magnitude smaller than the adiabatic correction, while in MoS$_2$ they are three orders of magnitude smaller and are practically zero.

Within the EF-DFPT, the first-order wavefunctions are given by   
\begin{equation} \label{eq:1storder_wfn}
\begin{split}
    \ket{\psi_{1}^{(1)}}=& \sum_{2\neq 1} \frac{\mel{\psi^{(0)}_{2}}{\hat{v}_{s}^{(1)}+\hat{v}_{geo}^{(1)}}{\psi^{(0)}_{1}}}{\epsilon^{(0)}_{1}-\epsilon^{(0)}_{2}} \ket{\psi^{(0)}_{2}} \\
    =& \sum_\lambda \sum_{2\neq 1} \frac{g_{n_2,n_1,\lambda}(\mathbf{k_1},\mathbf{q})}{\epsilon^{(0)}_{1}-\epsilon^{(0)}_{2}\pm \hbar\omega_{\mathbf{q}\lambda}} \frac{U_{\mathbf{q}\lambda}}{L_{\mathbf{q}\lambda}} \;\ket{\psi^{(0)}_{2}}\;.
\end{split}
\end{equation}
Fig.~\ref{fig:1}(d,e) shows the resulting modification of the electronic states through the overlap matrix, $S = \braket{\psi_{nk}^{(1)}}{\psi_{mk}^{(1)}}$, evaluated at momentum of the direct-gap in each system, namely the $C$ point in pentacene and the $K$ point in MoS$_2$. The wavefunctions remain normalized. Since the first order corrections are off-diagonal by construction the diagonal elements remain zero, while the off-diagonal elements quantify the EF-DFPT phonon-induced mixing between the unperturbed KS states. As expected, pentacene exhibits significantly stronger wavefunction mixing than MoS$_2$, reflecting enhanced electron–phonon coupling in the soft organic crystal, in good correspondence with the pronounced non-adiabatic effects observed in its dielectric response~\cite{cohen2024phonon}.

\begin{figure}[htbp]
    \centering
    \includegraphics[width=\linewidth]{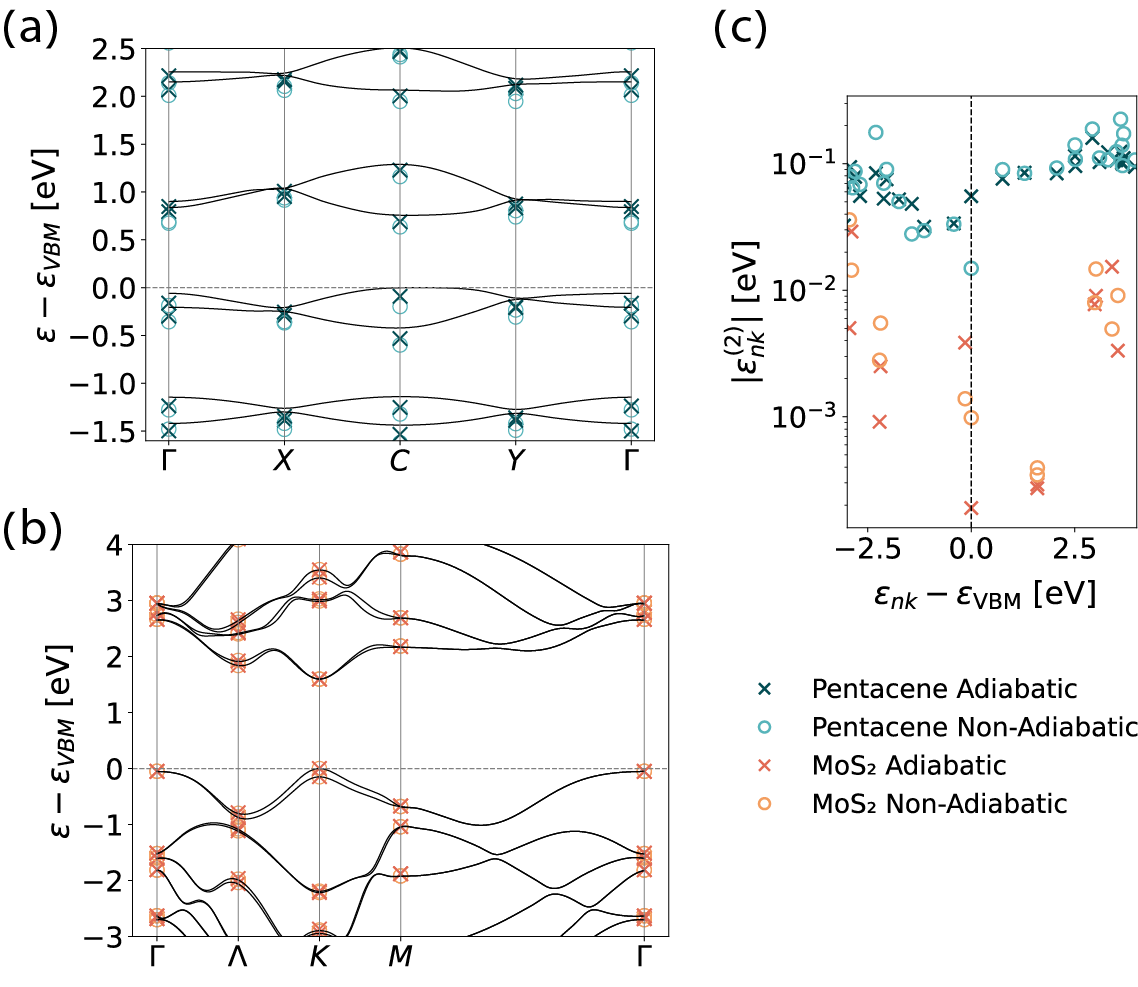}
    \caption{Second-order renormalized bandstructure for the pentacene crystal in blue (a) and monolayer MoS$_2$  in orange (b). Dark x symbols indicate the adiabatic corrected energies, and lighter 'o' symbols indicate the full EF-DFPT NA correction [Eq.\eqref{eq:eps2_U}. (c) shows the absolute value of the adiabatic and NA energy correction at the $C$ k-point in pentacene and the $K$ point in MoS$_2$. 
}
    \label{fig:2}
\end{figure}

Next, we turn to the changes in the electronic energies upon phonon interactions as captured in EF-DFPT. In conventional DFT calculations, bandstructure energies are evaluated for fixed nuclear positions within the BO approximation. When the electronic potential is expanded up to second order in the nuclear displacements 
$U_{\mathbf{q}\lambda}$, the electronic eigenenergies acquire corrections arising from both adiabatic and NA electron–phonon interactions. Within the EF framework, the second-order correction to the bandstructure energies is given by~\cite{cohen2025nonadiabaticity}
\begin{equation} \label{eq:eps2_U}
    \begin{split}
        &\varepsilon_{1}^{(2)} =\\ 
       &\sum_{2\neq 1} \sum_{\lambda}  \frac{\abs{g_{n_2,n_1,\lambda}(\mathbf{k_1},\mathbf{q})}^2}{\epsilon^{(0)}_{1}-\epsilon^{(0)}_{2}-\hbar\omega_{\mathbf{q}\lambda}}\bigg[ \frac{\abs{U_{\mathbf{q}\lambda}}^2}{L^2_{\mathbf{q}\lambda}}+ \frac{\hbar\omega_{\mathbf{q}\lambda}}{\epsilon^{(0)}_{1}-\epsilon^{(0)}_{2}-\hbar\omega_{\bar{\mathbf{q}}\lambda}}\bigg]\\
       &- \sum_{2\neq 1} \sum_{\lambda\lambda'}\frac{g^*_{n_2,n_1,\lambda'}(\mathbf{k_1},\mathbf{q})g_{n_2,n_1,\lambda}(\mathbf{k_1},\mathbf{q}) \hbar\omega_{\mathbf{q}\lambda}}{(\epsilon^{(0)}_{1}-\epsilon^{(0)}_{2}-\hbar\omega_{\bar{\mathbf{q}}\lambda'})(\epsilon^{(0)}_{1}-\epsilon^{(0)}_{2}-\hbar\omega_{\mathbf{q}\lambda})}\frac{{U_{\bar{\mathbf{q}}\lambda'}}}{L_{\bar{\mathbf{q}}\lambda'}}\frac{{U_{\mathbf{q}\lambda}}}{L_{\mathbf{q}\lambda}}\\
       &+\sum_{\mathbf{q}\lambda} g^{(2)}_{n_1,n_1;\lambda,\lambda}(\mathbf{k}_1,\mathbf{q},\bar{\mathbf{q}}) \frac{\abs{U_{\mathbf{q}\lambda}}^2}{L_{\mathbf{q}\lambda}^2}\;
    \end{split}
\end{equation}
where the last term includes the two-phonon Debye–Waller matrix elements $g^{(2)}_{n_1,n_1;\lambda,\lambda}(\mathbf{k}_1,\mathbf{q},\bar{\mathbf{q}})$.
Averaging over nuclear configurations using the nuclear probability density, the mixed term in the second line cancels. The resulting average second-order energy correction consists of two contributions: (i) a one-phonon term corresponding to the Fan–Migdal self-energy, including NA corrections through the phonon frequencies, and (ii) a two-phonon Debye–Waller term, in direct analogy with the standard DFPT formulation~\cite{giustino2017electron}
\begin{equation} \label{eq:eps2_avarage}
    \begin{split}
       \bar{\varepsilon}_{1}^{(2)} = \sum_{2\neq 1} \sum_{\lambda}  \frac{\abs{g_{n_2,n_1,\lambda}(\mathbf{k_1},\mathbf{q})}^2}{\varepsilon^{(0)}_{1}-\varepsilon^{(0)}_{2}-\hbar\omega_{\mathbf{q}\lambda}} 
       + \sum_{\mathbf{q}\lambda}
      g^{(2)}_{n_1,n_1;\lambda,\lambda}(\mathbf{k}_1,\mathbf{q},\bar{\mathbf{q}})\;.
    \end{split}
\end{equation}
Here, in contrast, the adiabatic second-order correction does not contain phonon frequencies in the energy denominator.
To compute these energy corrections in practice, we adopt the rigid-ion approximation and approximate the two-phonon Debye–Waller matrix elements as $g^{(2)}_{n_1,n_1;\lambda,\lambda}(\mathbf{k}_1,\mathbf{q},\bar{\mathbf{q}})=\abs{g_{n_1,n_1,\lambda}(\mathbf{k_1},\mathbf{q})}^2$ \cite{ponce2014verification, ponce2014temperature}. 
Fig.~\ref{fig:2}(a),(b) shows the adiabatic and NA corrections, resulting from the EF-DFPT calculation, to the bandstructure energies at high-symmetry points, as obtained from Eq.~\eqref{eq:eps2_U} for the cases of the pentacene crystal and monolayer MoS$_2$, respectively. Fig.~\ref{fig:2}(c) shows the value of the correction at the band gap, namely at the $C$ k-point in pentacene and the $K$ k-point in MoS$_2$. The values of both the adiabatic and NA corrections of MoS$_2$ are orders of magnitude smaller compared to pentacene.

Finally, we examine the impact of the non-adiabatic corrections on the static dielectric response, expected to be affected by the above-presented corrections in both the wavefunction states and the energy renormalization. 
The static inverse dielectric function, $\epsilon^{-1}$ ~\cite{hybertsen1986electron}, is  given by $\epsilon^{-1}_{GG'}(\mathbf{q}) = \delta_{GG'}-V(\mathbf{q}+\mathbf{G})\chi_{GG'}(\mathbf{q})$, where $V(\mathbf{q}+\mathbf{G})$ is the Coulomb interaction with the periodicity of reciprocal lattice vector $G$, and $\chi_{GG'}(\mathbf{q})$ is the static non-interacting  polarizability within the random-phase approximation (RPA). Within the EF-DFPT framework, the polarizability explicitly depends on the phonon-dressed electronic wavefunctions and the NA-corrected electronic energies, through
\begin{equation} \label{eq:chi}
\begin{split}
    &\chi_{GG'}(\mathbf{q})= \\&\sum^{occ}_2\sum^{unocc}_1 M_{21}(\mathbf{G})M^*_{21}(\mathbf{G}')\frac{1}{(\varepsilon_2^{(0)}+\varepsilon_2^{(2)})-(\varepsilon_1^{(0)}+\varepsilon_1^{(2)})}
\end{split}
\end{equation}
where $\varepsilon_i^{(2)}$ are the second-order corrections to the electronic energy level $i$, and the matrix elements
\begin{equation}
    M_{21}(\mathbf{G}) = \bra{\psi_2^{(0)}+\psi_2^{(1)}}e^{i(\mathbf{q}+\mathbf{G})\mathbf{r}}\ket{\psi_1^{(0)}+\psi_1^{(1)}}
\end{equation}
incorporate the calculated first-order phonon-induced corrections to the electronic wavefunctions, constructing the corrected NA phonon-dressed electronic states $\ket{\psi_i^{(0)}+\psi_i^{(1)}}$. Notably, non-adiabatic effects enter the dielectric response through both the numerator and denominator in Eq.~\eqref{eq:chi}.
\begin{figure}[htbp]
    \centering
    \includegraphics[width=\linewidth]{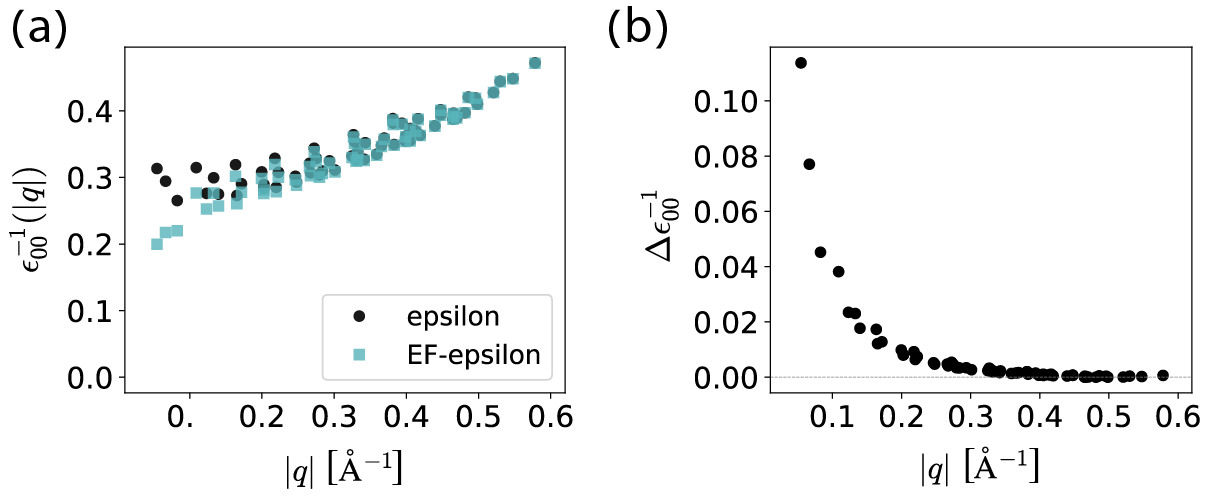}
    \caption{ (a) Dielectric response in the pentacene crystal from EF-DFPT, computed using EF-corrected wavefunctions (blue squares) and using ground state DFT wavefunctions (black circles). (b) The difference between the two cases shown in (a).   The EF-DFPT wavefunction corrections yield measurable modifications to the polarizability in the long-range interaction regime and, consequently, to the static dielectric matrix $\epsilon^{-1}_{GG'}$, demonstrating significant long-range NA effects on the electronic response in pentacene. 
 }
    \label{fig:epsilon}
\end{figure}

Fig.~\ref{fig:epsilon}(a) shows the head elements ($G=G'=0$) of the inverse dielectric function for the pentacene crystal computed following Eq.~\eqref{eq:chi}, using the EF-DFPT first-order corrected wavefunctions (blue squares) and, for comparison, using ground-state DFT wavefunctions within the BO adiabatic approximation (black circles). 
Fig.~\ref{fig:epsilon}(b) shows the difference between the two calculations. Our results point to a systematic reduction in the magnitude of the inverse dielectric function, mainly pronounced at the long-range interaction regime (small $\abs{q}$) where NA effects are included. This regime is associated with intermolecular interactions. This reduction thus corresponds to a phonon-induced enhancement of the dielectric screening, which was previously shown to be stronger at the long-range limit for the pentacene crystal~\cite{refaely2013gap}. These results are also consistent with the emergence of long-range interactions and weak-localization phenomena characteristic of Fr\"{o}lich-type polarons~\cite{dai2025polarons}.

The electronic dielectric function is a key quantity in introducing the self-consistent dependency of the electronic and nuclear degrees of freedom (as also stressed in the field-theoretic Hedin-Baym framework~\cite{giustino2017electron}). From this perspective, introducing phononic corrections, and in particular NA effects, into the dielectric response represents a further step toward a mutual and self-consistent treatment of the electron–phonon many-body system. The electron-phonon effects on the dielectric function are particularly relevant in materials with soft vibrational modes, where the static dielectric response is known to be strongly affected by phonon contributions~\cite{cockayne2000phonons}; our results for pentacene fall squarely within this class: the soft lattice and enhanced electron–phonon coupling lead to pronounced NA corrections to both electronic wavefunctions and energies, which in turn give rise to measurable modifications of the dielectric screening. By contrast, in monolayer MoS$_2$, characterized by stiff phonon modes and weaker electron–phonon coupling, NA effects are negligible, reinforcing the validity of the Born–Oppenheimer approximation for this material.

To summarize, our findings highlight the material-dependent nature of non-adiabatic effects and underscore the importance of identifying classes of materials in which NA physics plays a central role. The EF-DFPT approach is particularly powerful in this context, as it provides direct access to the microscopic ingredients, phonon-dressed orbitals, and energy renormalization, that introduce NA signatures. This method is not limited to dielectric screening but can be extended to other observables that depend sensitively on electronic structure, such as optical spectra, quasiparticle energies, charge transport, and excitonic properties. On the computational perspective, our work presents the first \textit{ab initio} implementation and application of EF-DFPT to realistic semiconducting solids. While the EF formalism has been extensively developed at the conceptual and model-system level, our results constitute the first quantitative assessment of non-adiabatic electron–phonon effects in solid-state materials. Methodologically, EF-DFPT offers a systematically improvable, perturbative route to incorporate NA effects within a Kohn–Sham-like framework that is compatible with standard electronic-structure workflows. Looking ahead, the present work lays the foundation for future developments, including self-consistent EF potentials, finite-temperature extensions, and applications to correlated and low-dimensional systems. More broadly, EF-based approaches hold significant promise for enabling predictive first-principles simulations of coupled electron–nuclear dynamics.

\begin{acknowledgments}
\vspace{-10pt}
Computational resources were provided by the ChemFarm local cluster at the Weizmann
Institute of Science. G.C. acknowledges an Institute for Environmental Sustainability (IES) Fellowship. 
E.K.U.G. acknowledges support from the Fondation de l'Ecole Polytechnique within the Gaspard Monge programme.  S.R-A acknoledges funding from the European Research Council (ERC), grant agreement No.ERC-2022-StG-101041159. 
\end{acknowledgments}
\maketitle

\end{document}